\documentclass[a4paper,10pt]{article}
\usepackage[utf8]{inputenc}
\usepackage{authblk}
\usepackage{setspace}
\usepackage[margin=1.25in]{geometry}
\usepackage{graphicx}
\graphicspath{ {./figures/} }
\usepackage{subcaption}
\usepackage{amsmath}
\usepackage{lineno}
\usepackage{xcolor}
\usepackage{hyperref}
\usepackage{braket}
\usepackage{cite}

\title{Investigating pump harmonics generation in a SNAIL-based Traveling Wave Parametric Amplifier}

\author[1,2]{A. Yu. Levochkina}
\author[1,2]{H. G. Ahmad}
\author[1,2]{P. Mastrovito}
\author[1,2]{I. Chatterjee}
\author[1]{G. Serpico} 
\author[1]{L. Di Palma} 
\author[1]{R. Ferroiuolo} 
\author[2]{R. Satariano}
\author[2]{P. Darvehi}
\author[3]{A. Ranadive}
\author[3]{G. Cappelli}
\author[3]{G. Le Gal}
\author[4]{L. Planat}
\author[1,2]{D. Montemurro} 
\author[5,2]{D. Massarotti}
\author[1,6]{F. Tafuri}
\author[3]{N. Roch}
\author[1,2]{G. P. Pepe} 
\author[2*]{M. Esposito}

\affil[1]{Department of Physics, University Federico II, Naples, 80126, Italy}
\affil[2]{CNR-SPIN, Complesso di Monte S. Angelo, via Cintia, Napoli, 80126, Italy}
\affil[3]{Universit\'e Grenoble Alpes, CNRS, Grenoble INP, Institut N\'eel, 38000 Grenoble, France}
\affil[4]{Silent Waves, Grenoble, France}
\affil[5]{Department of Electrical Engineering and Information Technology, University Federico II, Naples, 80125, Italy}
\affil[6]{CNR INO, Largo Enrico Fermi 6, Florence, 50125, Italy}
\affil[*]{martina.esposito@spin.cnr.it}

\date{\today}

\begin{document}
\maketitle

%%%%%% Abstract %%%%%%
\begin{abstract}
Traveling Wave Parametric Amplifiers (TWPAs) are extensively employed in experiments involving weak microwave signals for their highly desirable quantum-limited and broadband characteristics. However, TWPAs' broadband nature comes with the disadvantage of admitting the activation of spurious nonlinear processes, such as harmonics generation, that can potentially degrade  amplification performance. Here we experimentally investigate a Josephson TWPA device with SNAIL (Superconducting Nonlinear Asymmetric Inductive Element)-based unit cells focusing on the amplification behaviour along with the generation of second and third harmonics of the pump. By comparing experimental results with 
transient numerical simulations, we demonstrate the influence of Josephson junctions' fabrication imperfections 
on the occurrence of harmonics and on the gain behaviour. 
\end{abstract}

%%%%%% Main Text %%%%%%

%%%%%% Intoduction %%%%%%
\section{Introduction}
A Josephson TWPA (JTWPA) is in essence a nonlinear superconducting transmission line constituted of an array of Josephson junctions-based unit cells. In virtue of its quantum-limited noise property and wide bandwidth operation, such device is a highly valuable tool for microwave signal amplification \cite{White15, Bell15,  Macklin15, Zorin18, Miano19, Planat20, Ranadive22, Peng22, Braggio22, bartram23, Di_vora23, Roudsari23} and  entanglement generation \cite{Esposito22, Perelshtein22,Qiu23, Casariego_2023}. 
Amplification in JTWPAs occurs through a nonlinear wave-mixing process; when an intense pump at frequency $f_p$ and a weak signal at frequency $f_s$ are applied at the input of the device, the pump periodically modulates the nonlinear inductance of the Josephson junctions leading to an energy transfer from the pump itself to the signal mode and to an idler mode, at frequency $f_i$. Depending on the order of the nonlinearity, three wave-mixing ($f_p$ = $f_s$ + $f_i$) and four wave-mixing (2$f_p$ = $f_s$ + $f_i$) amplification processes can be distinguished \cite{Roy16,Aumentado20, Esposito21}. 
Due to the large TWPA bandwidth, energy from the pump can also leak creating undesired additional modes that still satisfy energy conservation, such as sidebands (combinations of pump and signal/idler tones) \cite{Sakuraba63, Mckinstrie05, Chaudhuri15, Erickson17, Peng22} and pump harmonics \cite{Obrien14, Dixon20, Nilsson23}. 
The dominant spurious tones are the second and the third harmonics of the pump, which in turn depend on the nonlinear coefficients associated with three wave-mixing (3WM) and four wave-mixing (4WM) interactions respectively. 

In this article, we explore the behaviour of pump harmonics arising in JTWPAs, focusing on a \textit{Reversed Kerr TWPA} \cite{Ranadive22} device,  characterized by a SNAIL (Superconducting Nonlinear Asymmetric Inductive Element)-based unit cell \cite{Frattini18}. 
A key advantage of adopting a SNAIL-based JTWPA is the possibility to control both 3WM and 4WM nonlinear coefficients by an external magnetic flux \cite{Ranadive22, Roudsari23, Katayama23}, 
making it an ideal testing device for investigating spurious modes. %the generation of pump harmonics. 
In the specific case of a \textit{Reversed Kerr TWPA},
3WM nonlinear processes are inhibited by deliberately designing the unit cells with alternating flux polarity (since the 3WM nonlinear coefficient is an odd function of the flux) \cite{Ranadive22}. In such a way, in an ideal device, the second harmonic generation (SHG) of the pump, at frequency $f_{\text{SH}} = 2 f_p$, should be in principle negligible at any flux value,
while the third harmonic generation (THG), at frequency $f_{\text{TH}} = 3 f_p $, is expected to be dominant and flux dependent.

We present an experimental study of gain, second, and third harmonics generation as a function of magnetic flux and pump amplitude.
In contrast to ideal predictions, we  observe a non-negligible presence of pump second harmonic generation. By means of numerical simulations, performed with \textit{WRspice} \cite{WRSPICE} software, we can reproduce the experimental behaviour when a spread in the critical current of the Josephson junctions (JJs) is taken into account. 

Understanding the effects of fabrication tolerance on JTWPAs' performance is an open challenge and has recently been a topic of in-depth numerical investigations \cite{Peatain23, Kissling23}. 
Our detailed study of nonlinear processes in the presence of JJs fabrication imperfections contributes to explaining the origin of unexpected harmonics in JTWPAs, providing new experimental insights supported by robust numerical models.

%%%%%% Device Description %%%%%%
\section{Device description}
\begin{figure}[htbp]
    \centering
    \includegraphics[width=0.85\textwidth]{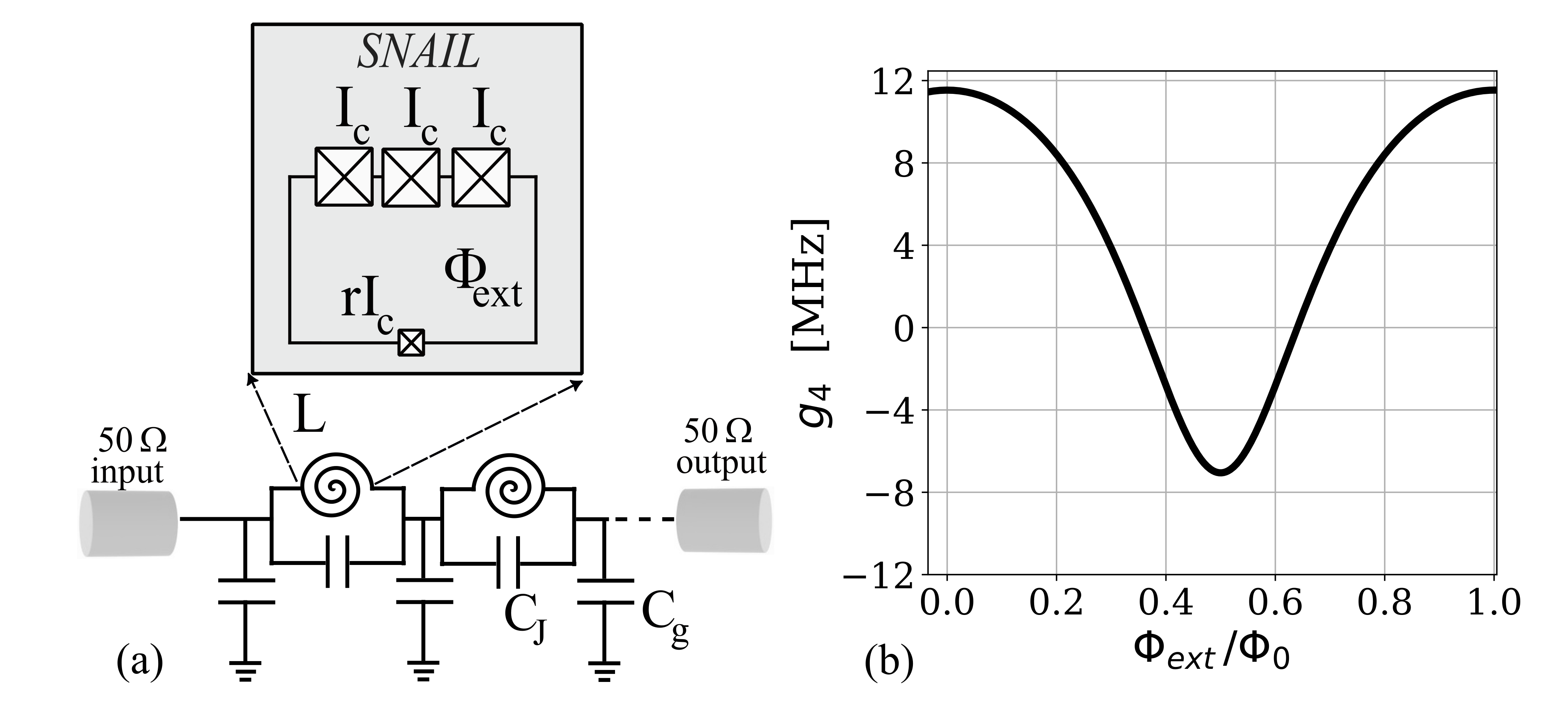}
    \caption{(a) Sketch of the \textit{Reversed Kerr TWPA} device under investigation. $C_g$ indicates the unit cell capacitance to ground, $C_J$ the unit cell Josephson capacitance and $L$ the inductance per unit cell. The SNAIL symbols are shown with opposite orientation for adjacent cells indicating the alternating flux polarity design. The inset is a sketch of the SNAIL element which includes three large JJs with critical current $I_c$ and one small JJ with critical current $rI_c$. The main device's parameters and the methods for their estimation are reported in Appendix. (b) 4WM nonlinear coefficient $g_4$ as a function of applied magnetic flux for the device under study. }
    \label{fig:1}
\end{figure}

The \textit{Reversed Kerr TWPA} under investigation is nominally identical to the one reported in \cite{Ranadive22} and it is sketched in Fig. \ref{fig:1} (a). In the figure, $I_C$ indicates the critical current, $C_g$ indicates the unit cell capacitance to ground, $C_J$ the unit cell Josephson capacitance and $L$ the inductance per unit cell. Additional details on circuit parameters are reported in Appendix. 
A key design aspect for this device is that SNAILs are oriented in such a way that the external magnetic flux has opposite sign for adjacent unit cells (alternating flux polarity), aiming at the suppression of the overall 3WM nonlinearity \cite{Zorin21} and therefore of SHG. 
By assuming negligible 3WM nonlinear processes, 
the wave equation describing the propagation through the device can be written as 
\begin{equation} \label{eq:1}
    \frac{\partial^2 \phi}{\partial x^2} - \frac{1}{\omega_0^2}\frac{\partial^2 \phi}{\partial t^2} +
    \frac{1}{\omega_J^2}\frac{\partial^4 \phi}{\partial x^2 \partial t^2} - 
    8 g_4 \frac{R_Q}{\pi \omega_0 Z}\frac{\partial }{\partial x}\left[{\left(\frac{\partial \phi }{\partial x}\right)}^3\right]
    = 0 \, ,
\end{equation}
where $\omega_J = \sqrt{1/(L C_J)}$ and $\omega_0 = \sqrt{1/(L C_g)}$ are the plasma frequency and the characteristic frequency of the transmission line, $R_Q = h/(4e^2)$,  $Z = \sqrt{L/C_g}$ is the characteristic impedance of the transmission line and $g_4$ is the flux-dependent 4WM nonlinear coefficient \cite{Ranadive22} indicating the rate at which the 4WM nonlinear processes occur. The dependence of $g_4$ on the external magnetic flux is reported in Fig \ref{fig:1} (b), while the full analytical expression is reported in Appendix.

\begin{figure}[htbp]
    \centering    \includegraphics[width=1.0\textwidth]{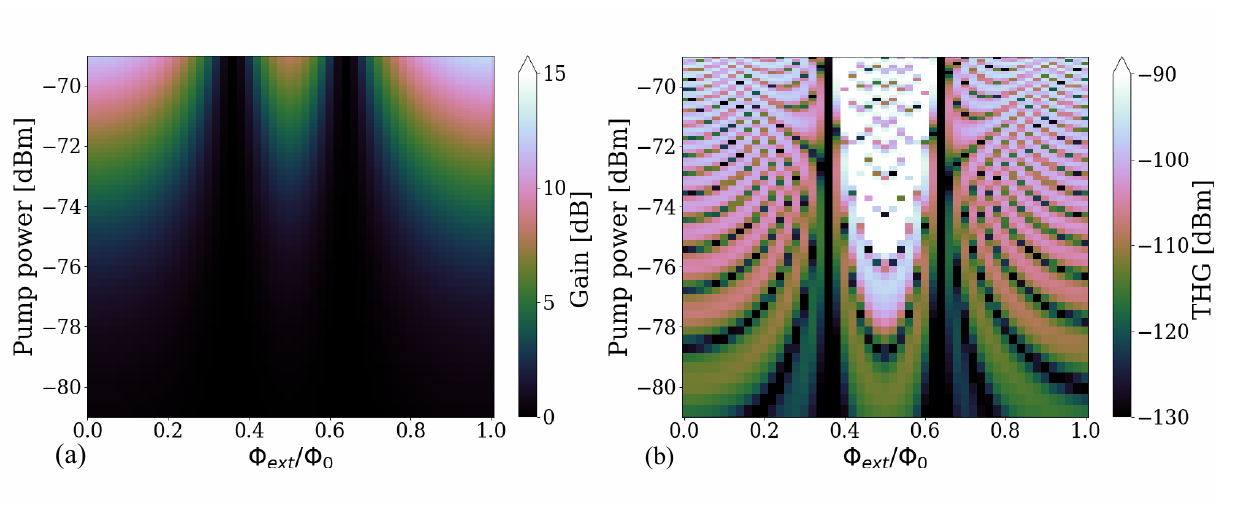}
    \caption {Couple Mode Equation (CME) theory
     predictions for gain (a) and THG of the pump (b) as a function of pump power and external magnetic  flux. Pump Pump frequency $f_p = 4 \, \text{GHz}$, signal frequency $f_s = 3.9 \,  \text{GHz}$. The signal power is assumed to be lower than 1 dB compression point power of the device. The plots have been obtained  with the open source software tool CMETANA (Couple Mode Equation Theory Analyzer) \cite{CMETANA}  which provides a user-friendly environment for CME predictions of JTWPAs starting from unit cells parameters. The exact analytical expressions are reported in Appendix and in CMETANA’s documentation. } 
    \label{fig:2}
\end{figure}

Following the standard approach developed in \cite{Yaakobi13,Obrien14} for JTWPAs and subsequently generalized to the case of SNAIL-based unit cells \cite{Bell15}, we can use the standard Couple Mode Equation (CME) theory to calculate the expected 4WM gain and the amplitude of the THG of the pump. Specifically, when 4WM gain is calculated, pump, signal and idler modes are included in the wave equation ansatz while, for the THG case, only the pump and its third harmonic are considered. 
The analytically calculated CME predictions are reported in Fig. \ref{fig:2} as a function of pump power and external magnetic flux. In order to accommodate all the relevant modes within the typical 4-12 GHz experimental frequency region of circuit-QED \cite{Blais21}, we set the pump frequency $f_p = 4 \, \text{GHz}$ and the signal frequency $f_s = 3.9 \,  \text{GHz}$. 
The exact analytical expressions are reported in Appendix. 

Being both 4WM processes, the gain and THG display the same flux-dependent behaviour associated with the typical 4WM $g_4$ nonlinear coefficient (Fig. \ref{fig:1} (b)). 
Analytical CME predictions provide a useful benchmark when a new device is experimentally tested for the first time. Nevertheless, as reported in the following section, this simplified theoretical approach is not suitable to describe the full broadband experimental behaviour due to the constraint of including only a limited number of propagating tones.

%%%%%% Experimental Results %%%%%%
\section{Experimental results}
\begin{figure}[htbp]
    \centering
    \includegraphics[width=1.0\textwidth]{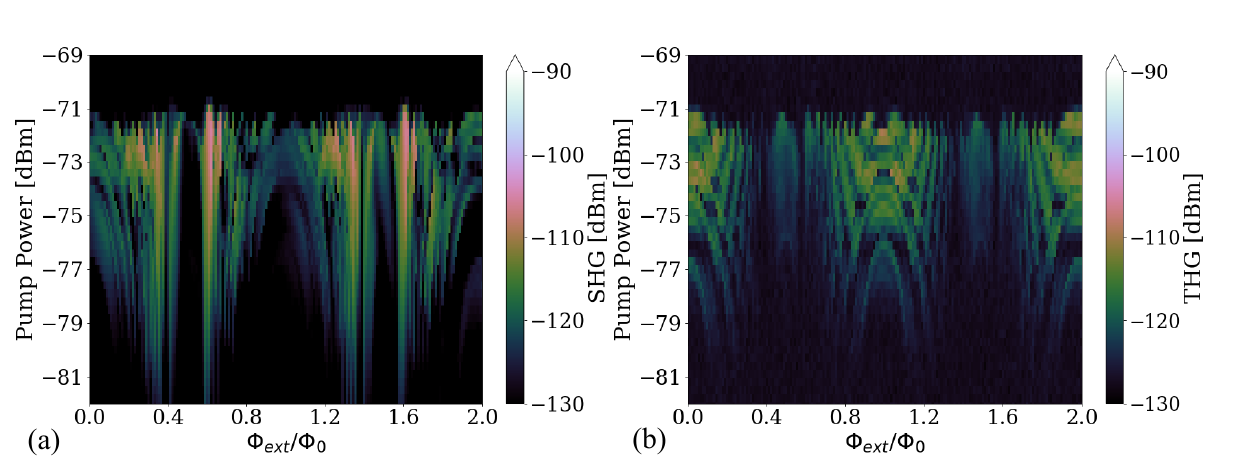}
    \caption{Experimental study of harmonics generation.  (a)  SHG and (b) THG of the pump measured with a Spectrum Analyzer as a function of pump power and applied flux. Pump frequency $f_p = 4 \, \text{GHz}$, signal frequency $f_s = 3.9 \,  \text{GHz}$, signal power at device input $P_s = -110 \,  \text{dBm}.$   
    }
    \label{fig:4}
\end{figure}

%%%%%%%%%%%%%%%%%%%%%%%%%%%%%%%%%%%%%%%%%%%%%%%%%%%%%%%%%%%%%%%%%%%%%%%
The device has been experimentally measured at a temperature of 10 mK in a dilution refrigerator with a standard microwave experimental setup (see Appendix). 
\begin{figure}[htbp]
    \centering
    \includegraphics[width=1.0\textwidth]{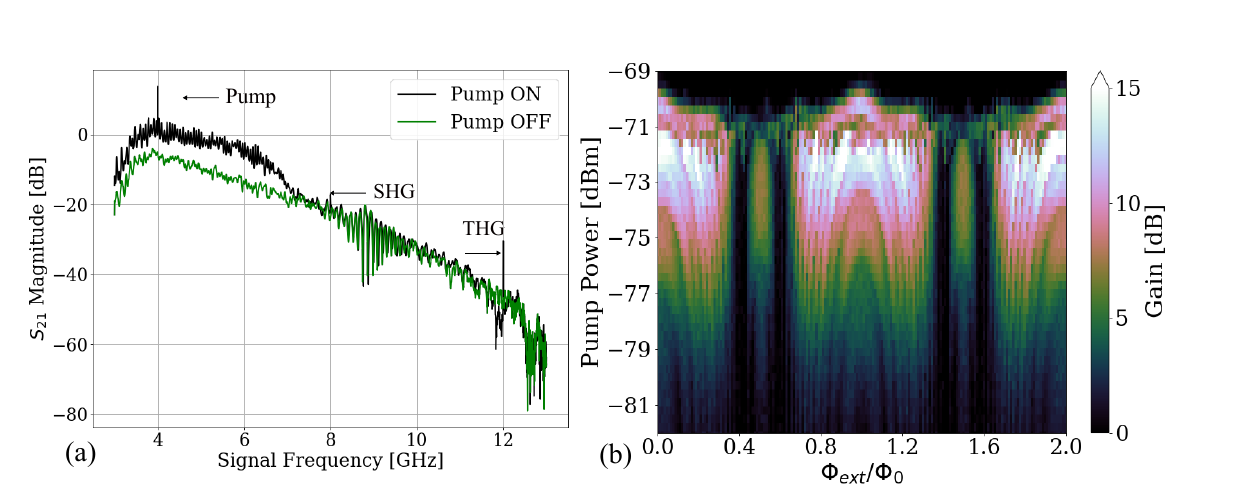}
    \caption{Gain measurements. (a) VNA measurements of the uncalibrated transmission, $S_{21}$ traces, for pump OFF and ON; external magnetic flux $\Phi_{\text{ext}}/\Phi_{0} = 0.5$. Pump power at the device is $P_{p} = -74 \,  \text{dBm}$. (b) SA gain measurement (difference between pump on and pump off of the power spectral density at the signal frequency) as a function of pump power and external magnetic flux.  Pump frequency $f_p = 4 \, \text{GHz}$, signal frequency $f_s = 3.9 \,  \text{GHz}$, signal power at device input $P_s = -110\,  \text{dBm}$.   }  
    \label{fig:3}
\end{figure}
In Fig. \ref{fig:3} (a), we report the device transmission ($S_{21}$ parameter) measured with a Vector Network Analyzer (VNA) with and without the presence of the pump tone at frequency $f_p = 4 \, \text{GHz}$, 
when external magnetic flux 
$\Phi_{\text{ext}}/\Phi_{0}$ is equal to 0.5. 
We deliberately chose the VNA settings in order to be able to detect the pump. When the pump is on, we observe 4WM gain along with the generation of the pump's harmonics. The VNA plots have an illustrative purpose, providing an example of the frequency-dependent behaviour of the device when pumped. 
To accurately investigate the generation of 4WM gain in the presence of pump harmonics, we perform a detailed spectral power characterization of the relevant frequency tones using a spectrum analyzer (SA). We set the pump frequency $f_p=4 \, \text{GHz}$ and we measure the intensity of a signal tone, at frequency $f_s=3.9 \,  \text{GHz}$, as a function of the external magnetic flux and pump power (\ref{fig:3} (b)). In the same conditions, we also measure the SHG, at frequency $f_{\text{SH}} = 8 \,  \text{GHz}$, and the THG of the pump, at frequency $f_{\text{TH}} = 12 \,  \text{GHz}$. The measured power is calibrated at the output of the device. The signal power at device input is $P_s = -110 \,  \text{dBm}$. 
The experimental results of harmonics measurements are reported in Fig. \ref{fig:4}.

We observe 4WM gain (\ref{fig:3} (b)) and THG (\ref{fig:4} (b)) which periodically change with the applied magnetic flux, following the behaviour of the SNAIL 4WM nonlinear coefficient, as qualitatively expected from CME theory (Fig. \ref{fig:1} (b)).
In contrast with what is ideally predicted for a SNAIL JTWPA with alternated flux polarity structure, 
we observe a non-negligible flux depended SHG power spectral density which for some flux and pump power values is comparable or larger than the THG one. We stress that for an accurate quantitative  comparison between the measured spectral power of SHG and THG one should also consider that losses in the device under study increase for larger frequencies, specifically they approximately increase of 1 dB per GHz \cite{Ranadive22}. In addition, when the pump input power at the device exceeds roughly -71 dBm, we measure a sudden drop in the device's experimental transmission due to the saturation of the Josephson junction critical current which is not described by CME theory.

Finally, we also verify that the presence of the signal tone at the JTWPA input does not significantly influence the pump's harmonics generation (see Appendix).

\section{Numerical simulations}
\begin{figure}[htbp]
    \centering
    \includegraphics[width=0.85\textwidth]{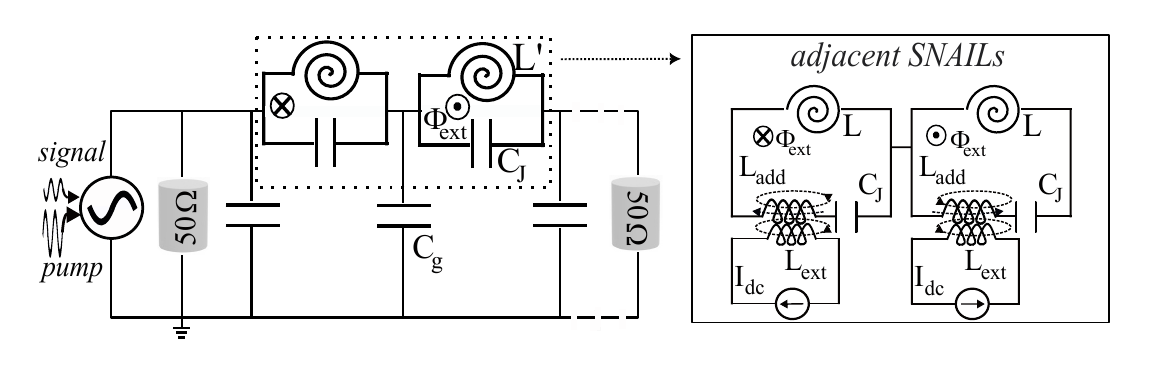}
    \caption{Equivalent circuit of a SNAIL TWPA adopted for numerical simulations in the \textit{WRspice} environment.  Here $L_{add}$ is an inductance added to the SNAIL for enabling flux-biasing simulation. $C_J$ is the unit cell Josephson capacitance and $L$ is the inductance per unit cell. $C_g$ indicates the unit cell capacitance to ground. $L_{ext}$ is the inductance of the additional loop. $I_{dc}$ is associated with the WRspice direct current source element. Inversed SNAILs orientation is reproduced by alternating current direction in the DC source.}
    \label{fig:6}
\end{figure}

To capture the origin of the observed experimental behaviour we perform \textit{WRspice} numerical transient simulations by adopting the equivalent circuit sketched in Fig. \ref{fig:6} and neglecting losses. 
To model the flux tunability, we add a small linear inductance, $L_{add} = 1 pH$ in each SNAIL element and consider an auxiliary external inducting loop with inductance $L_{ext} = 1 uH$ \cite{aggarwal21, Levochkina24IEEE}. In this way, the flux is coupled via the external inductance and is controlled by a simulated external source of direct current.

Fabrication-related imperfections among different unit cells  are also included in the numerical simulations. 
Specifically, we introduce a spread in the critical currents of all Josephson junctions in the full device (700 cells) by assuming a random Gaussian distribution with mean values equal to the ones estimated from fabrication parameters and linear experimental characterization of the device (see Appendix). The standard deviation of the random Gaussian distribution varies from $\delta/\braket{I} = 0$ (ideal case in which all junctions with the same designed size have identical critical current) to $\delta/\braket{I}  = 15 \%$ (typical upper limit for Josephson junctions' arrays fabrication tolerance \cite{Peatain23,Kissling23}).

\begin{figure}[htbp]
    \centering
    \includegraphics[width=1.0\textwidth]{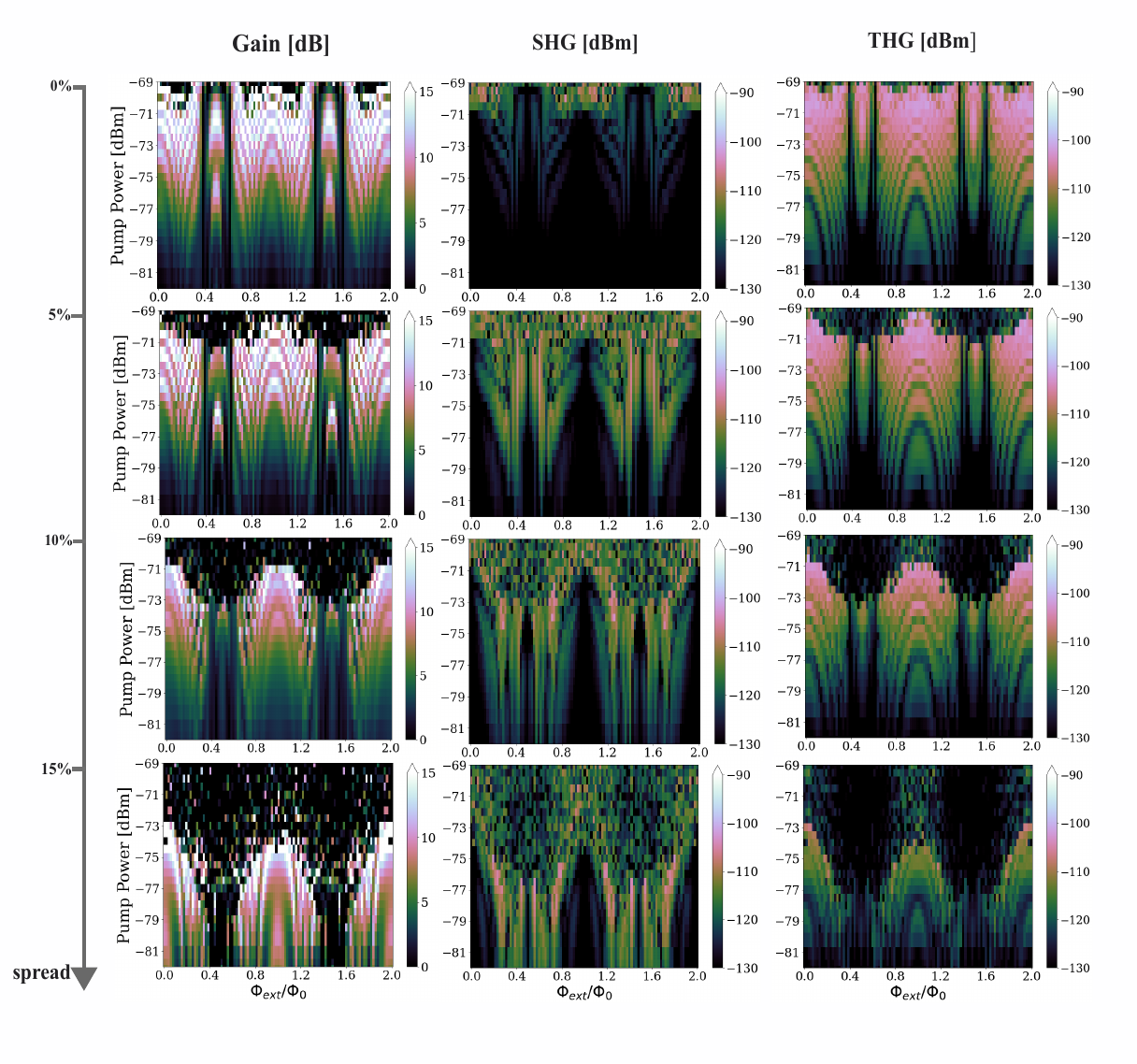}
    \caption{\textit{WRspice} simulation results. The first column is Gain, the second column is SHG and the third column is THG of the pump as a function of pump power and external magnetic flux with different values of JJ critical currents' parameters spread. The spread value changes from $\delta/\braket{I} = 0 \%$ (top row) up to $\delta/\braket{I} = 15 \%$ (bottom row). Pump frequency $f_p = 4 \, \text{GHz}$, signal frequency $f_s = 3.9 \,  \text{GHz}$, signal power at device input $P_s = -110 \,  \text{dBm}$.}
    \label{fig:7}
\end{figure}

In Fig. \ref{fig:7}, we show the results of the transient simulations by plotting the signal gain (\textit{pump on - pump off}, differential output spectral power at $f_s$) and the output spectral power at the pump's harmonics frequencies as a function of pump power and external magnetic flux. 

The simulated gain and THG are periodic with the external flux: the observed maxima and minima correspond to the experimentally obtained flux points for all the tested spread values. On the other hand, we find a relevant impact of the JJ's critical current spread on the behaviour of gain and THG as a function of pump power, 
observing a decrease of the pump saturation power, from roughly $-70 \, \text{dBm}$ to roughly $-75 \, \text{dBm}$ when the spread amount increases from $0$ to $15 \%$.
In addition, we also find that the spread in JJ critical currents affects the absolute maximum amplification value, going from about $15 \, \text{dB}$ for $\delta/\braket{I} = 0$ to $12 \, \text{dB}$ for $\delta/\braket{I}  = 15 \%$. 
We interpret the simultaneous reduction of the maximum values for the two 4WM processes, gain and THG, as an effect of the reduced saturation power. By increasing the spread, some of the junctions have a lower critical current, bringing the full device to saturate at lower pump power. Therefore, the maximum achievable gain and THG amplitude is reduced.

Finally, an important outcome following from the numerical simulations is the influence of the spread of JJ's critical currents on the generation of the second harmonic of the pump.  While for an ideal device (no spread in the critical current values) SHG of the pump is almost fully suppressed for pump powers below the saturation of the critical current (about -70 dB), its intensity increases as the spread percentage increases, showing the best qualitative agreement with the flux-dependent experimental behaviour for typical values of fabrication-related JJ's critical current spread of $5\%$.

We interpret the increase of the SHG with increasing spread in the Josephson junctions’ critical currents as an effect of the increasing disorder in the alternating flux polarity design. 
The latter ideally guarantees a full cancellation of the 3WM nonlinearity in the approximation of wavelengths much larger than the unit cells \cite{Zorin21}. This is valid because the 3WM coefficient $\beta$ associated with each SNAIL (see Eq. \ref{eq:4} in Appendix) is an odd function of the external flux. 
In the case of  0 \% spread, adjacent SNAILs are identical but with opposite flux polarity. This means that the corresponding $\beta$ coefficient will have identical absolute value but opposite sign, resulting in an overall suppression of 3WM processes like SHG.
By switching on the spread in junction critical currents, adjacent SNAILs and the corresponding $\beta$ amplitude are not identical anymore, and their difference gets bigger as the spread in junction critical current increases, thus increasing the overall $\beta$ and so the efficiency of SHG.
The residual SHG at 0\% spread below the saturation power is attributed to a deviation from the assumption of fully cancelled 3WM nonlinearity for the alternated flux polarity design, corresponding to the approximation of wavelengths much larger than the unit cells \cite{Zorin21}.

\begin{figure}[htbp]
    \centering
    \includegraphics[width=1.0\textwidth]{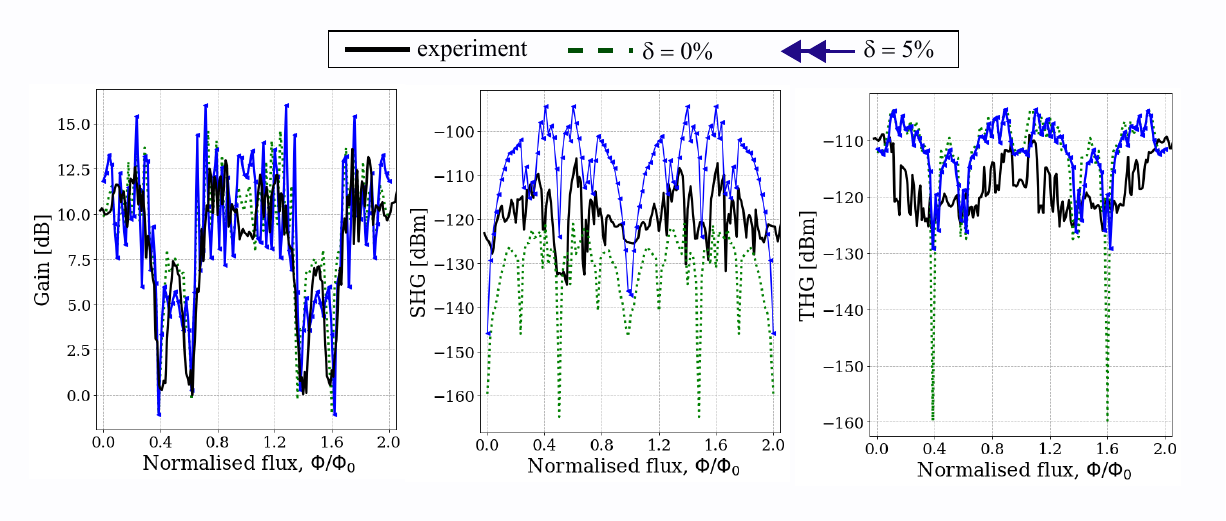}
    \caption{Horizontal cuts of experimentally  and numerically obtained 2d-color-maps for   $-74 \, \text{dBm}$ pump power. Simulations are reported for the case of  $0\%$ and $5\%$ JJs'critical current spread.}
    \label{fig:8}
\end{figure}

For a more accurate comparison with the experimental results, in Fig. \ref{fig:8}, we report horizontal cuts of both experimentally and numerically obtained 2d-color-maps for a selected pump power of $-74 \, \text{dBm}$, highlighting the simulations results for $0\%$ and $5\%$ spread.
 %%%
%%%%
In Fig \ref{fig:8}, we observe qualitative agreement between experiments and theory for gain and SHG, where the experimental curve lies in between the simulated ones with 0\% and 5\% spread. However, for the THG case, the simulations are both overestimating the power observed experimentally. We attribute this discrepancy to the fact that our model does not include losses, which in the real device are considerably higher at the frequency of the THG than at the SHG and signal frequencies \cite{Ranadive22}.  The inclusion of losses in the simulations is a promising perspective for future investigations in order to improve the agreement between simulations and experiments at higher frequencies. 

\section{Conclusion}
In conclusion, we presented an experimental and numerical study of harmonics generation in a prototypical JTWPA with SNAIL-based unit cells. Our findings demonstrate the impact of Josephson junctions fabrication imperfections on the generation of the pump's harmonics and on gain performance. 
By reporting for the first time a detailed experimental investigation of the dominant pump harmonics during the operation of a JTWPA as an amplifier, our work provides new insights for understanding the origin of spurious tones in such devices, fostering the development of both mitigation strategies and potential new exploitation methods of higher order processes in JTWPAs.

\section*{Data availability statement}
The experimental data that support the findings of this work are openly available in the Zenodo repository: 10.5281/zenodo.11498844
\section*{Acknowledgements}
This work is supported by the European Union under Horizon Europe 2021-2027 Framework Programme Grant Agreement no. 101080152 and under Horizon 2020 Research and Innovation Programme Grant Agreement no. 731473 and 101017733; and by PNRR MUR project PE0000023-NQSTI. This project has received funding from the European Union’s Horizon 2020 research and innovation programme under grant agreement no. 899561.

%%%%%%%%%%%%%%%%%%%%%%%%%%%%%%%%%%%%%%%%%%%%%%%%%%%%%%%%%%%%%%%%%%%%%%%%%%%%%%%%
%start of the appendix
\newpage
\section*{Appendices}
\appendix
%\appendixautorefname{}
%\renewcommand{\appendixpage}{Appendix}

%%%%%%%%%%%%%%%%%%%%%%%%%%%%%%%%%%%%%%%%%%%%%%%%
\section{Experiment setup}
\begin{figure}[htpb]
    \centering    \includegraphics[width=1.0\textwidth]{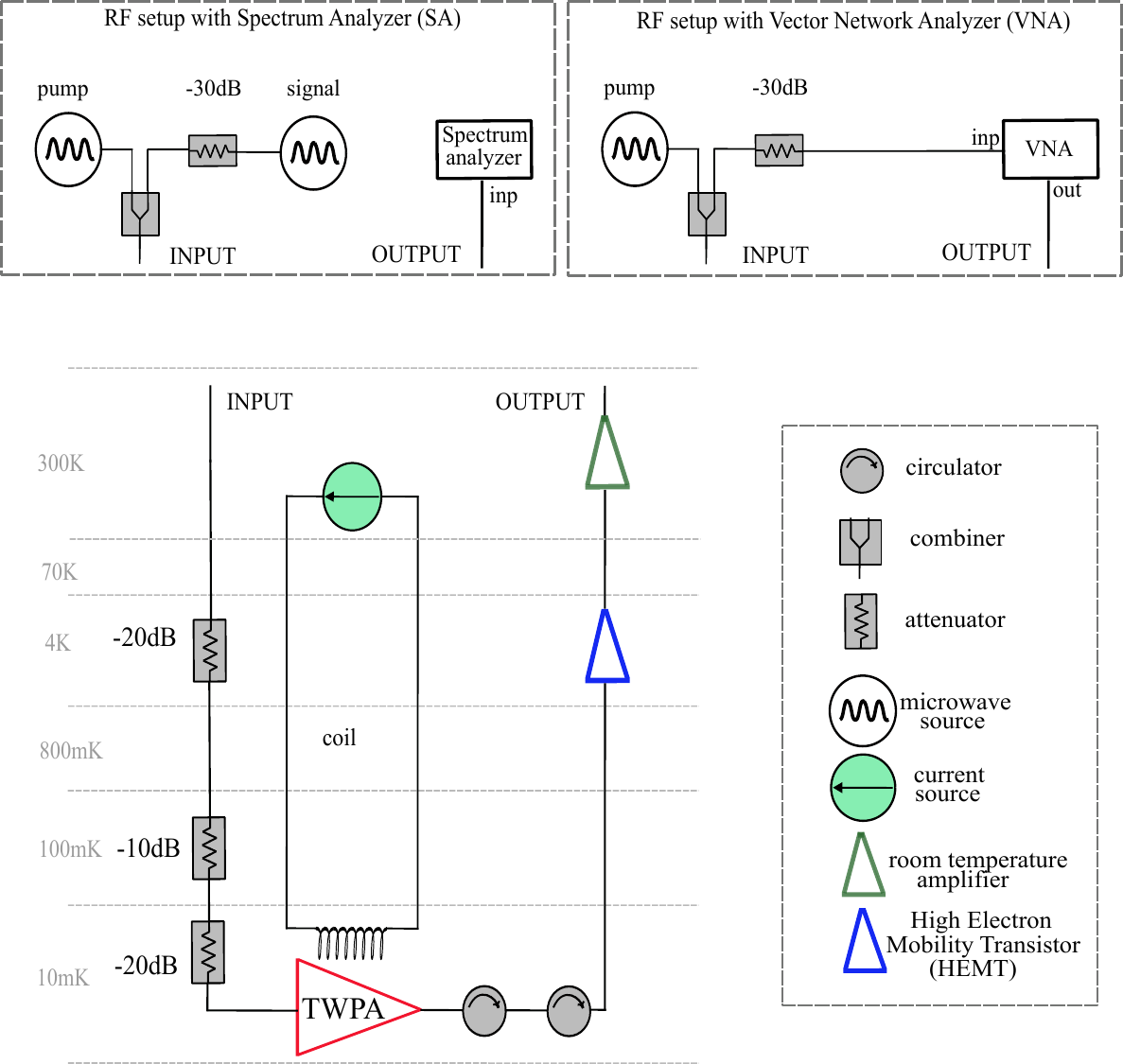}
    \caption{Sketch of the experimental setup. The top insets represent the room-temperature part of the setup showing the two adopted configurations using either the Spectrum Analyzer or the  Vector Network Analyzer. 
    The bottom scheme reports the cryogenic setup inside the dilution refrigerator. }
    \label{fig:10}
\end{figure}

The cryogenic part of the experimental setup is shown in Fig. \ref{fig:10}.
The total input line attenuation is estimated considering the sum of all the cryogenic attenuators, 
%The input line includes cryogenic attenuators 
reaching in total 
-50 dB, and the nominal  attenuation of the input RF cables (-10 dB). The output line includes two circulators in series at the 10 mK stage of the dilution refrigerator and a High Electron Mobility Transistor (HEMT) cryogenic amplifier at the 4 K stage. The HEMT working frequency range is from 4 GHz to 12 GHz with a nominal amplification of about 40 dB. At 300 K, a room-temperature amplifier providing about 30 dB amplification is installed. For flux biasing the TWPA, a superconducting coil is placed alongside the device holder.

%We performed different types of TWPA analysis, since we use 
We adopt two main room-temperature setups (Fig. \ref{fig:10} insets). For transmission measurements like the one reported in Fig. \ref{fig:3} (a), a VNA (Vector Network Analyzer) is used, and its output is combined with the one of an RF source used for pumping the device. 
For spectral power characterizations like the one reported in Fig. \ref{fig:3} (b) and Fig. \ref{fig:4}, a SA (Spectrum Analyzer) is used and two different RF sources are combined to send the pump and signal in input. 

\section{Characterizations of the device in the linear regime}
\label{Appendix_Linear}
We perform a preliminary linear measurement of the device under investigation in order to characterize the flux tunability and estimate the critical current $I_c$ and the SNAIL ratio $r$.
Other circuit parameters, such as the ground capacitance $C_g$, the SNAIL Josephson capacitance $C_J$, and the number of cells, are obtained from design and fabrication values \cite{Ranadive22} and reported in Table \ref{tab:1}.
\begin{table}[ht]
    \caption{Parameters of the device extracted from design and fabrication values.}   
    \centering 
    \begin{tabular}{ll}
            \hline
            \textbf{Parameter} & \textbf{Value}  \\  
            \hline
            number of cells, $N$ & $700$ \\ 
            
            Josephson capacitance, $C_J$ & $31$  $fF$ \\
            ground capacitance, $C_g$ & $250$ $fF$\\
            \hline
            \end{tabular}
    \label{tab:1}
\end{table}
\\
The linear characterization is performed following the procedure reported in the Supplementary Information of reference \cite{Ranadive22}.
\begin{figure}[htbp]
    \centering
    \includegraphics[width=1.0\textwidth]{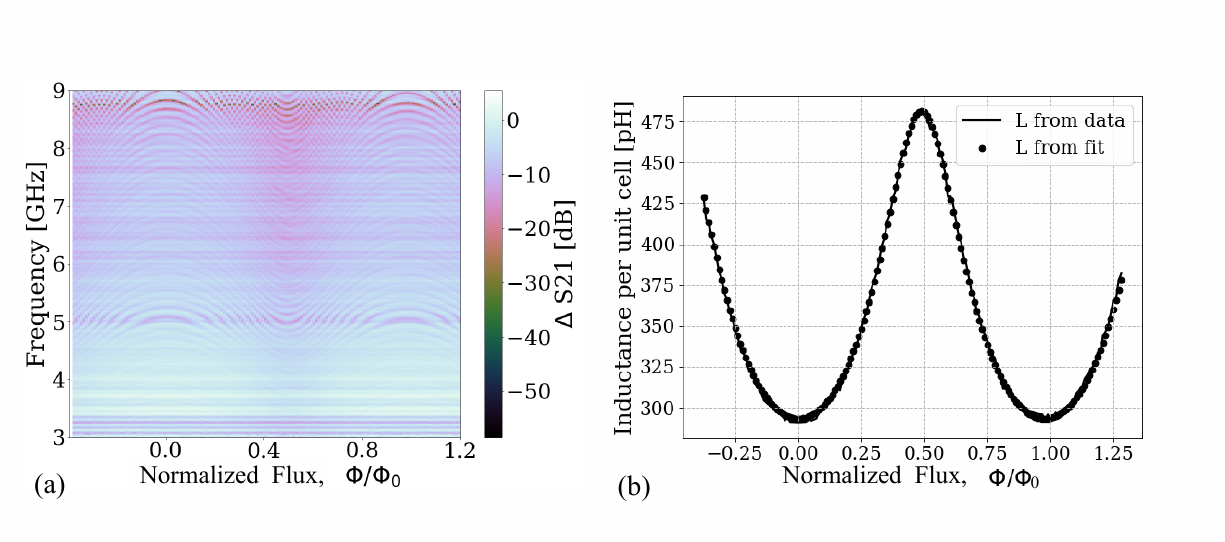}
    \caption{Linear characterisation and parameters extraction (a) Difference between transmissions of TWPA and "dummy device". (b) Experimentally obtained and fitted inductance of unit cell.}
    \label{fig:11}
\end{figure}
In detail, we sweep the current in the coil and measure $S_{21}$ TWPA transmission  (magnitude and phase) as a function of the external magnetic flux using the VNA setup with a low VNA signal power and no pump in input. The periodic modulation of the TWPA transmission with the applied magnetic flux is due to the flux-dependence of the characteristic impedance, $Z = \sqrt{L/C_g}$.
By subtracting from the $S_{21}$ phase a "reference phase" obtained by substituting the TWPA with a "dummy PCB device", we get the net transmitted phase $\Theta_{\text{TWPA}}(\omega)$, which is related to the dispersion relation as follow $k(\omega)=\Theta_{\text{TWPA}}(\omega)/N$, where N is the number of unit cells. By assuming disperion without mixing, $k(\omega)$ = $N {\sqrt{L C_g}\omega}/{\sqrt{1-\frac{\omega^2}{L C_J}}}$, for each value of the magnetic flux we can extract L (inductance per unit cell). The latter values are reported in Fig. \ref{fig:11} (b) together with the fit function that takes into account the analytic expression for L as a function of the critical current value $I_c$ and the SNAIL ratio $r$ which are used as fitting parameters.From the best fit we estimate $I_c = 2.6 \, \mu A$ and $r = 0.08$.

Fig \ref{fig:11} (a) reports the device losses as a function of frequency and applied flux.

%%%%%%%%%%%%%%%%%%%%%%%%%%%%%%%%%%%%%%%%%%%%%%%%%%%%%%%%%%%%
\section{Influence of input signal on pump harmonics generation }
\begin{figure}[htb!]
    \centering   
    \includegraphics[width=0.9\textwidth]{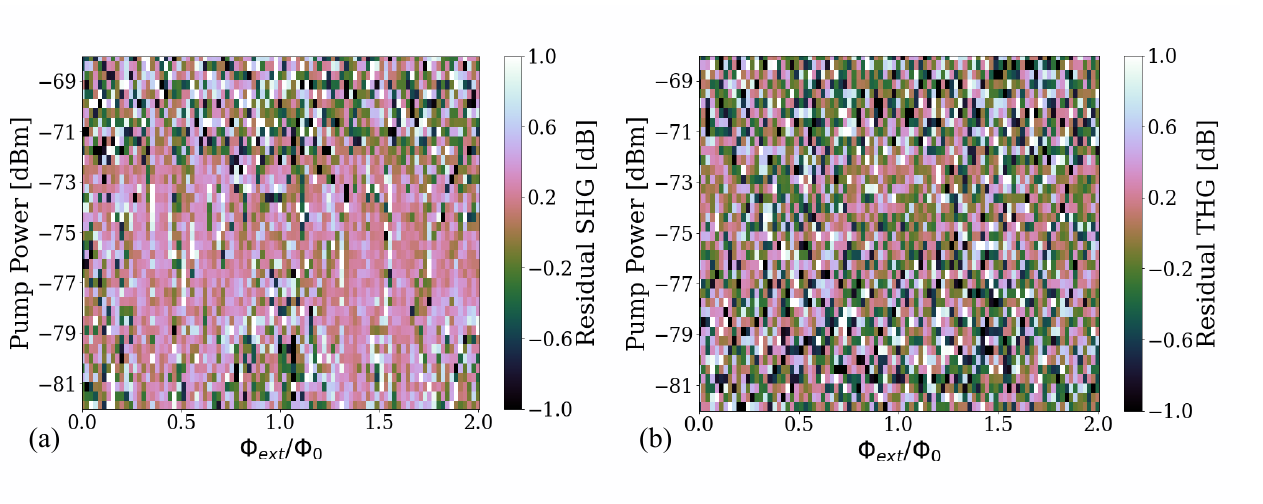}
    \caption{  
    {Residual pump harmonics spectral power obtained performing the difference between the case with \textit{input signal on} and \textit{input signal off} for SHG (left) and THG (right). Pump frequency $f_p = 4 \, \text{GHz}$, signal frequency $f_s = 3.9 \,  \text{GHz}$, signal power at device input $P_s = -110 \,  \text{dBm}.$  }
    }
    \label{residual}
\end{figure}

We provide the measurement results of the SHG and THG with and without the input signal. In Fig. \ref{residual} we plot the experimental residual spectral power for SHG and THG obtained calculating the difference between the case with \textit{input signal on} and with \textit{input signal off}. We observe that the presence of the signal does not significantly influence the pump's harmonics generation (see Appendix).

%%%%%%%%%%%%%%%%%%%%%%%%%%%%%%%%%%%%%%%%%%%%%%%%%%%%%%%%%%%
\section{Experiments in a large range of external magnetic flux}
\begin{figure}[htb!]
    \centering   
    \includegraphics[width=1.0\textwidth]{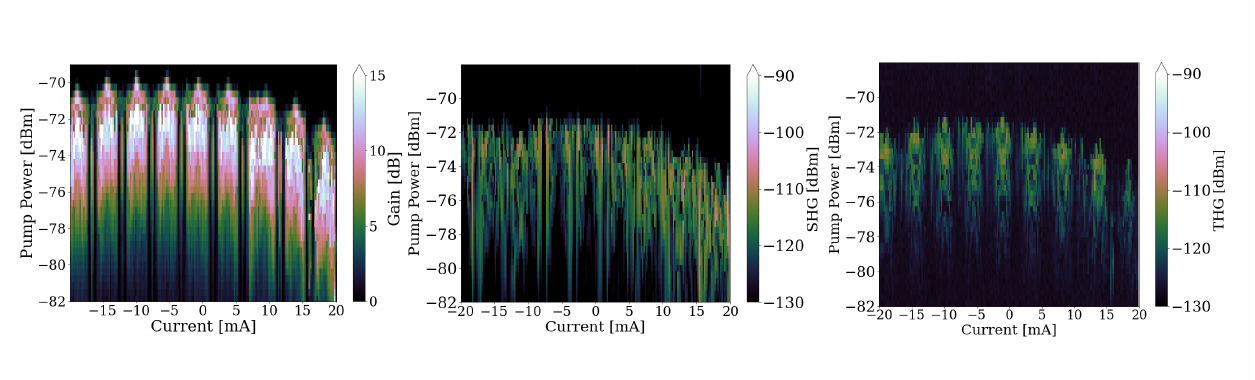}
    \caption{  
    {Experimental behaviour in a large range of external magnetic flux. Gain (left),  SHG (middle) and  THG (right) as a function of pump power and current in the coil placed 	underneath the sample. Pump frequency $f_p = 4 \, \text{GHz}$, signal frequency $f_s = 3.9 \,  \text{GHz}$, signal power at device input $P_s = -110 \,  \text{dBm}.$  }
    }
    \label{fig:13}
\end{figure}

In Fig. \ref{fig:13}, we report the experimental results for TWPA  SHG and THG in an extended range of external magnetic flux that here we express in terms of the current in the coil. 
The reported range is the maximum range of current values before starting to observe an increase in the base cryostat temperature. Along with the expected flux-dependent periodicity, we also observe an additional slow modulation with the external flux. (The flux range for the data reported in Fig. \ref{fig:3} and Fig. \ref{fig:4} corresponds to the range between -10 mA and 0 mA, in terms of direct current of the coil.) 

The observed slow modulation with the external flux can be explained by the fact that during the experiment no magnetic shielding has been adopted, making the measurements sensitive to additional environmental magnetic fields. Specifically, as recently reported in \cite{Mukhopadhyay2023, Kuzmin23}, the Josephson junction critical current can be affected by magnetic field lines along the insulating layer of Josephson junctions due to the "bending" of the total perpendicular magnetic field. 

\section{Coupled mode equations theory} 

Here we report the main derivation steps to obtain the analytical expressions for gain and THG displayed in Fig. \ref{fig:2}, following the standard coupled mode equations approach developed in \cite{Yaakobi13,Obrien14, Bell15}. First we provide the definition for the main flux dependent parameters. To do so, we start from the Taylor expansion of the current $I_L$ through a SNAIL unit cell:
\begin{equation} \label{eq:2}
    \frac{I_L(\phi^* + \phi)}{I_C} \approx 
    \Tilde{\alpha} \phi - \Tilde{\beta}(\phi)^2 - \Tilde{\gamma}(\phi)^3,
\end{equation}
where $I_c$ is the Josephson junction critical current, and the expansion is about a flux $\phi^*$ such that $I_L(\phi ^*)=0$. The coefficient in Eq \ref{eq:2} are defined as follows:
\begin{equation} \label{eq:3}
    \Tilde{\alpha} = r \cos \phi^* + \frac{1}{3}\cos\left(\frac{\phi^*-\phi_{ext}}{3}\right),
\end{equation}
\begin{equation} \label{eq:4}
    \Tilde{\beta} = \frac{1}{2}\left[r \sin \phi^* + \frac{1}{9} \sin\left(\frac{\phi^*-\phi_{ext}}{3}\right)\right],
\end{equation}
\begin{equation} \label{eq:5}
    \Tilde{\gamma} = \frac{1}{6}\left[ r \cos\phi^* + \frac{1}{27} \cos\left(\frac{\phi^*-\phi_{ext}}{3}\right)\right] \, .
\end{equation}
Considering the alternating flux polarity design of the device under investigation, we neglect the 3WM nonlinear coefficient, assuming $\Tilde{\beta} = 0$ for any value of the external flux. Note that the coefficients $\Tilde{\gamma}$ and $\Tilde{\alpha}$ are associated with the $g_4$ 4WM coupling rate via the following relation: $\hbar g_4 = \frac{\Tilde{\gamma}}{2\Tilde{\alpha}}E_C$, where $E_C = e^2/(2 C_g)$ is the charging energy.

Given the definitions above, eq. \ref{eq:1} can be rewritten as
\begin{equation} \label{eq:7}
    \frac{\partial^2 \phi}{\partial x^2} - \frac{1}{\omega_0^2}\frac{\partial^2 \phi}{\partial t^2} +
    \frac{1}{\omega_J^2}\frac{\partial^4 \phi}{\partial x^2 \partial t^2} - 
    \gamma\frac{\partial }{\partial x}\left[{\left(\frac{\partial \phi }{\partial x}\right)}^3\right]
    = 0 \, ,
\end{equation}
where $\gamma = \widetilde\gamma /\widetilde{\alpha}$, $L = \Phi_0/(2 \pi I_C \Tilde{\alpha})$ and $\Phi_0$ being the magnetic flux quantum.

To find the analytical expression for the 4WM gain, the following ansatz is considered \cite{Yaakobi13,Obrien14}:
 \begin{equation} \label{eq:8}
    \phi(x,t)=\frac{1}{2}\left[A_p(x)e^{i(k_px - w_pt)}+ A_s(x)e^{i(k_sx - w_st)}+
    A_i(x)e^{(k_ix - w_it)}+ c.c.
    \right] .
\end{equation}
By inserting the ansatz solution in Eq. \ref{eq:7},  one obtains a set of three coupled differential equations describing the propagation of pump, signal and idler tones. Using a number of approximations, such as the slowly varying envelope approximation, the approximation of uniform transmission line, undepleted pump assumption and the assumption of a large pump amplitude relative to signal and idler, the final expression for the power gain is:
\begin{equation} \label{eq:13}
    G_{\text{power}} = \cosh^2(gx) + \frac{\Delta k^2}{4g^2}\sinh^2(gx) \, ,
\end{equation}
where g is the reduced gain coefficient given by
\begin{equation}
 g =    \sqrt{\left(\frac{k_s^2 \, k_i^2 \, (2k_p - k_i) \, (2 k_p - k_s) \omega_p^4}{k_p^6 \, \omega_s^2 \, \omega_i^2}\right) \alpha_p^2 - {\left(\frac{\Delta k}{2}\right)}^2} \, .
\end{equation}
The total phase mismatch is defined as $\Delta k = k_s + k_i - 2 k_p + 2\alpha_p - \alpha_i - \alpha_s$, where the coefficients $\alpha_{\textrm{m}}$ ($\textrm{m} = p, s, i$) are defined as follows:
\begin{equation}
    \alpha_p = \frac{3 \, \gamma \, \omega_0^2 \, k_p^5}{8 \, \omega_p^2} {\left| A_p \right|}^2 \quad \textrm{,} \quad \alpha_s = \frac{3 \, \gamma \, \omega_0^2 \, k_p^2 \, k_s^3}{4 \, \omega_s^2} {\left| A_p \right|}^2 \quad \textrm{,} \quad \alpha_i =\frac{3 \, \gamma \, \omega_0^2 \, k_p^2 \, k_i^3}{4 \, \omega_i^2} {\left| A_p \right|}^2 \, ,
\end{equation}
and the wave-vectors for signal, idler and pump can be obtained from the chromatic dispersion relation 
\begin{equation}
     k = \frac{\omega }{\omega_0\sqrt{1 - \omega^2/ \omega_J^2}}.
     \label{linear_dispersion2}
\end{equation}

A similar approach can be adopted to obtain the analytical expression for the THG amplitude \cite{Obrien14}. In this case, the ansatz solution includes only pump  and the third harmonic waves:
\begin{equation} \label{eq:14}
    \phi(x,t)=\frac{1}{2}\left[A_p(x)e^{i(k_px - w_pt)}+ A_h(x)e^{i(k_hx - w_ht)}+ c.c.
    \right] ,
\end{equation}
where $\omega_h  = 3 \omega_p$. The final expression for THG amplitude is given by:
 %%%%%%%%%%%%%%%%%%%%%%%%%%%
 \begin{equation}
A_{h} =   \kappa_2 \frac{1 - e^{i x (\Delta k + 3\kappa_0 - \kappa_1)}}{\Delta k + 3\kappa_0 - \kappa_1} \,  e^{-i \kappa_1 x}
     \label{sol_th},
 \end{equation}
with $\Delta k = 3 k_p - k_h$ being the chromatic phase mismatch, and the coefficients defined as follows, 
  \begin{equation}
 \kappa_ 0 = \gamma \frac{3 \, k_p^5 \, \omega_0^2 \, |A_{p0}|^2}{8 \, \omega_p^2}, \qquad    \kappa_1 = \gamma \frac{6 \, k_p^2 k_h^3 \, \omega_0^2 \, |A_{p0}|^2}{8 \, \omega_p^2},  \qquad \kappa_2 = \gamma \frac{3 \, k_p^4 k_h \, \omega_0^2 \, A_{p0}^3}{8 \, \omega_p^2} \, ,
 \end{equation}
with $|A_{p0}|$ the pump amplitude at the input of the device. 

%%%%%%%%%%%%%%%%%%%%%%%%%%%%%%%%%%%%%%%%%%%%%%%%%%%%%%%%%%%%%%%%%%%%%
\section{Frequency-dependent gain profile}
\begin{figure}[htb]
    \centering
    \includegraphics[width=0.70\textwidth]{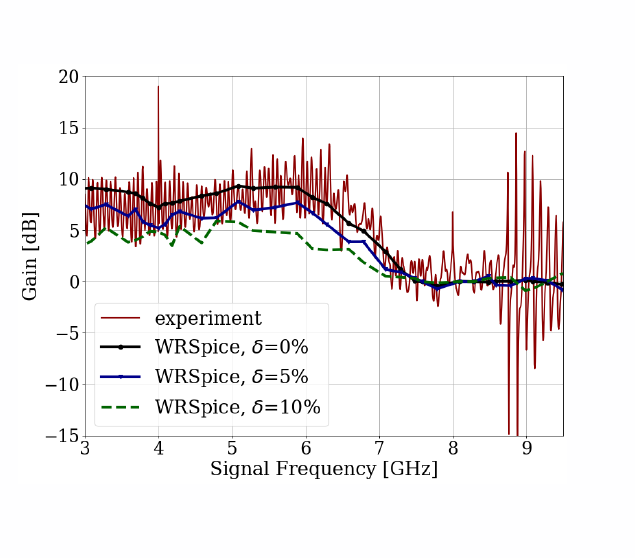}
     \caption{Comparison between the experimental frequency dependent gain profile (differential trace of the “pump ON” and “pump OFF” traces reported in Fig. \ref{fig:3} (a)) and WRSpice simulations for spread values $\delta = 0\%$,  $\delta = 5\%$ and $\delta = 10\%$. External magnetic flux $\Phi_{\text{ext}}/\Phi_{0} = 0.5$. Pump power at the device is $P_{p} = -74 \,  \text{dBm}$.}
    \label{fig:14}
\end{figure}

%%%%% Citations in the text %%%%%%
In Fig. \ref{fig:14} we report a comparison between the experimental frequency-dependent gain profile (differential trace of the “pump ON” and “pump OFF” traces in Fig. \ref{fig:3} (a)) and simulations for different spread values of the Josephson junctions critical current. 
For low frequencies, lower than about $4.5 
  \, \text{GHz}$, we observe qualitative agreement between the experimental data and the simulations for typical values of fabrication-related JJ’s critical current spread of $5\%$. This is consistent with the results reported in the main text for a representative signal frequency at $3.9 
 \, \text{GHz}$. The agreement between experiments and $\delta = 5\%$ simulations gets slightly worse for higher frequencies. This could be explained considering the decrease of dielectric losses with increasing power due to the saturation of two-level systems (TLS) in the real device \cite{Pappas11, McRae20}. This effect could lead to an overestimation of the experimental gain for higher frequencies, where losses are higher \cite{Ranadive22}.

%%%%%%%%%%%%%%%%%%%%%%%%%%%%%%%%%%%%%%%%%%%%%%%%%%%%%%%%%%%%
\section{SNAIL systematic fabrication imperfections influence}
In this section we explain in detail how in our simulations we took into account systematic fabrication imperfections associated with the single unit cell.
\begin{figure}[htb]
    \centering
    \includegraphics[width=0.70\textwidth]{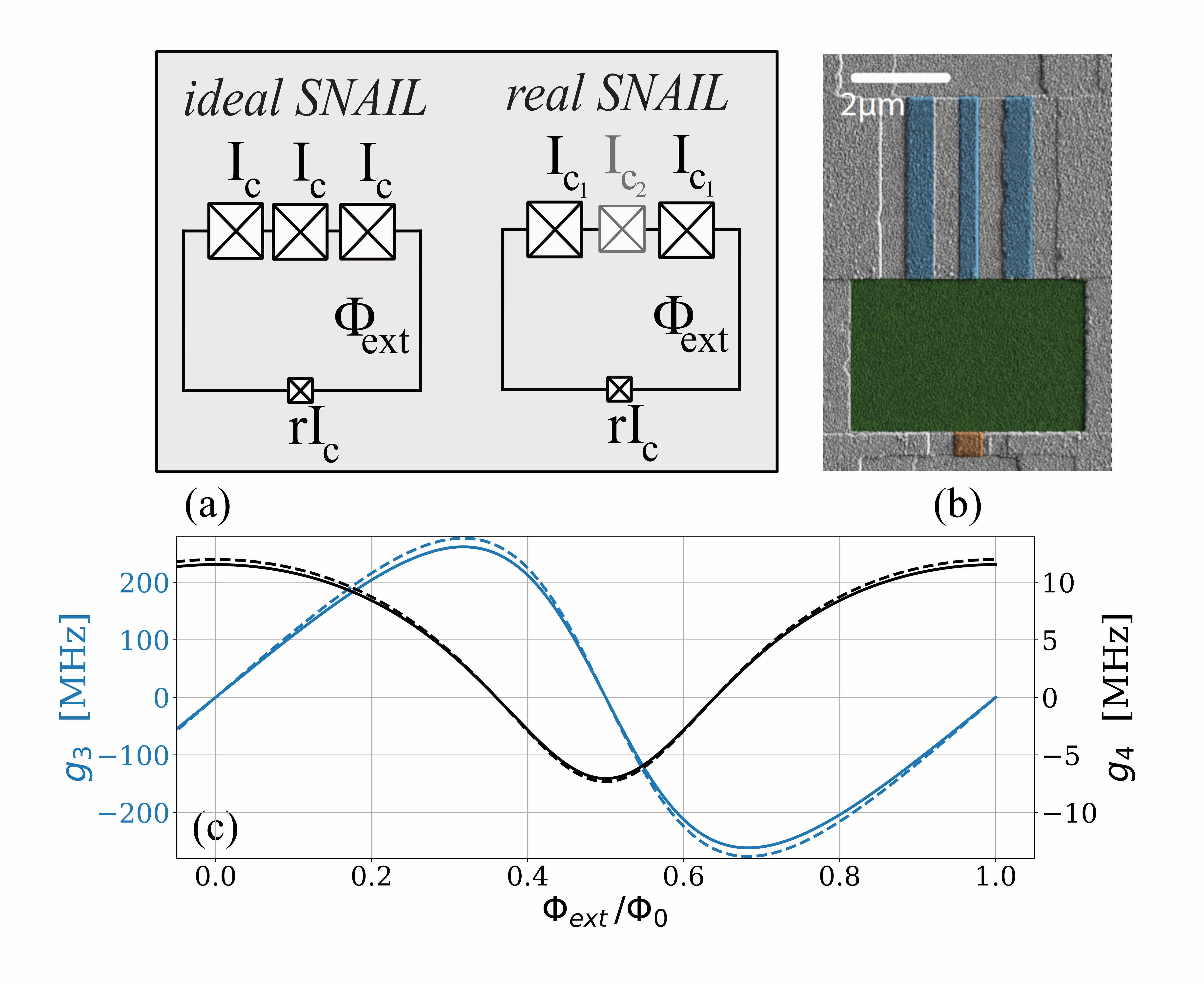}
     \caption{SNAIL fabrication imperfections influence. (a) Sketches of an ideal and a real SNAIL. (b) SEM SNAIL picture adapted from \cite{Ranadive22}. (c) 3WM ($g_3$) and 4WM ($g_4$) non-linear coefficients as a function of flux for both ideal (continuous lines) and real (dashed lines) SNAIL unit cell.}
    \label{fig:12}
\end{figure}
%
%\subsection{Device unit cells}
Each SNAIL in the investigated device consists of three large Josephson junctions (with critical currents $I_c$) and one small junction (with critical current $rI_c$). 
The three big JJs are all identical by design (sketch in Fig. \ref{fig:12} (a) left inset), however, the Dolan bridge JJ fabrication method for the SNAIL typically induces a dimension variation such that the junction located in the middle turns out slightly smaller than the side ones (see sketch in Fig. \ref{fig:12} (a) right inset). Such fabrication imperfection can be appreciated in Fig. \ref{fig:12} (b), which illustrates a scanning electron microscopy (SEM) image, adapted from \cite{Ranadive22}, of a typical SNAIL in the device. In order to investigate the effect of such systematic fab imperfection, from the SEM image we estimated the ratio between the area of the two big top junctions and the area of the smaller middle top junction $a_1/a_2 = 1.3$. In our simulations, we took into account this effect assuming that the average critical current for the three top junctions is given by the value $I_c$ estimated in Appendix \ref{Appendix_Linear}. 

In addition, for a better comprehension of the effect of such systematic imperfection, we compared the analytical predictions for 3WM and 4WM nonlinear coefficients $g_3$ and $g_4$ of an ideal SNAIL and the ones of a real SNAIL (modelled as in the "real SNAIL" sketch in Fig. \ref{fig:12} (b) - inset) obtained using the recently developed python package "NINA" (Nonlinear Inductive Network Analyzer) \cite{Miano23}, which allows to exactly compute the Taylor expansion coefficients of the effective potential energy function of an arbitrary flux-biased superconducting loop.
The result of such comparison is reported in Fig. \ref{fig:12} (c), which shows $g_3$ and $g_4$ coefficients as a function of the external magnetic flux for both an ideal (continuous lines) and a real SNAIL (dashed lines). We observe a negligible difference in the two cases and conclude that the aforementioned systematic fabrication  imperfection is not significantly affecting the nonlinear processes investigated in this work.

\newpage{}
\bibliographystyle{unsrturl}
\bibliography{sample}

\end{document}